\documentclass[12pt]{article}

\usepackage{amsmath,cite,hyperref}

\begin{document}

%%%%%%%%%%%%%%%%%%%%%%%%%%%%%%%%%%%%%%%%%%%%%%%%%%%%%%%%%%%%%%%%%%%%%

\begin{titlepage}

\title{Two-field constant roll inflation} 
\author{{\normalsize Andrei Micu} \\[.5cm]
   {\it\normalsize Department of Theoretical Physics,}\\
   {\it\normalsize Horia Hulubei National Institute for Physics and
     Nuclear Engineering}\\
   {\normalsize Str. Reactorului nr. 30, RO--077125, M\u{a}gurele,
     Romania} \\
   {\normalsize E-mail: \texttt{amicu@theory.nipne.ro}}}
\date{}

\maketitle

\begin{abstract}\noindent

Starting from the idea of realising constant roll inflation in string
theory we develop the constant roll formalism for two scalar fields.
We derive the two-field potential which is compatible with a
constant roll regime and discuss possible applications to string-models.

\end{abstract}

\vfill

\thispagestyle{empty}

\end{titlepage}

%%%%%%%%%%%%%%%%%%%%%%%%%%%%%%%%%%%%%%%%%%%%%%%%%%%%%%%%%%%%%%%%%%%%%%%%%%%%

\section{Introduction}

It is almost out of doubt that inflation has to be part of any
cosmological model. While from the GR point of view only the energy
content of the universe is relevant, from a particle physics point of
view there is a big question of how the conditions for inflation are
realised. As string / supergravity models offer most promising
perspectives to unify particle physics and gravity a lot of effort has
been spent in order to accommodate inflation in such theories. The
simplest idea is to generate inflation from a cosmological constant 
which in turn is given by the vacuum energy.
Even if, thanks to supersymmetry, Minkowski and AdS
solutions are quite easy to find \cite{CHSW,dWSHD,DPN}, dS solutions,
which should be relevant for inflation, 
are completely forbidden in many cases \cite{MN}. More recently, a series
of conjectures regarding the consistency of a theory coupled to
quantum gravity, known as the swampland criteria, have been formulated 
\cite{BCV,DvR,OOSV}. It was immediately realised that these criteria may be in
direct conflict with conditions for inflation
\cite{AOSV}.\footnote{Note that as long as the precise details for
  inflation are not known many inflationary options in agreement with
  the swampland criteria can be found \cite{Das}.}
It is therefore important to find models of inflation which are not in
the swampland according to these criteria.

In recent years a new type of inflation has been put forward
\cite{MS1} which was dubbed constant roll inflation. 
The novelty is that imposing a constant roll regime for the inflaton,
the exact background solution can be written down explicitly, while
the parameter defining the constant roll need not be in any
sense \emph{small} unlike in the slow roll case. This may ensure that
the constant roll condition is safe with regard of the swampland criteria
and moreover since the constant roll parameter is directly related to the
$\eta$ parameter defined in slow roll inflation, this new type of
inflation may constitute a simple workaround to the well known
supergravity $\eta$-problem \cite{DRT}. It is therefore natural to
analyse whether constant roll inflation can be realised in string theory.
One aspect to keep in mind is that for a single inflaton field, the
constant roll potential which was derived in \cite{MS1,ASW}, turns out
to be completely fixed and it is not clear whether
concrete string models can accommodate such potentials. 
A more relevant question to be asked is under what conditions
multi-field constant roll inflation can take place, since models
derived from string theory usually involve many scalar fields. 
In this note we shall give an answer to this question for a model with
two scalar fields.

The outline of the paper is the following.
We begin with a short review of single field constant roll solution in
Section 2 and then try to implement the corresponding potential in a
simple string model in Section 3. We shall see that the string
potential comes quite close to the form required by the constant
roll conditions, but nonetheless the identification of the string
potential with the constant roll one is still questionable. This will
motivate us to study in detail the two-field constant roll regime in
Section 4. We
derive the explicit solutions for the fields and for the scalar
potential and note that the shortcomings of the single field potential
may have a resolution in this more general case.

\section{Constant roll inflation}

In this section we shall briefly review constant roll conditions and
the solution for the case of a single scalar field. This was first
discussed in \cite{MS1} and further studied in more details in
\cite{MS2,OO}. Other developments appeared in \cite{NOO,MS3,CMP,OOS}.
A systematic study of constant roll inflation appeared
more recently \cite{ASW} and this reveals more possibilities then
originally proposed in \cite{MS1}. In the following we shall closely 
follow \cite{ASW}.

The matter action just contains the kinetic term for the inflaton and
a scalar potential
\begin{equation}
  \label{mat1f}
  S_{matter} = \int \sqrt{-g} d^4 x \Big( - \frac12 \partial_\mu
  \phi \partial^\mu \phi - V( \phi) \Big) \; , 
\end{equation}
and we use the ``mostly plus'' convention for the Minkowski
metric. In a spatially flat FRW background where the metric is given
by 
\begin{equation}
  d s^2 = -dt^2 + a(t)^2 d \vec x^2 \; , 
\end{equation}
the equations of motion read
\begin{eqnarray}
  \label{Eom1f}
  3 H^2 & = & \frac12 \dot \phi^2 + V \; ; \nonumber \\
  -2 \dot H & = & \dot \phi^2 \; ; \\
  0 & = & \ddot \phi + 3H \dot \phi + \frac{\delta V}{\delta \phi} \; ,\nonumber
\end{eqnarray}
where $H$ is the Hubble parameter and is given as $H=\frac{\dot a}{a}$.
It is important to notice that for the case of a single scalar field, once the
Einstein equations are fulfilled the equation of motion for the scalar
field is identically satisfied. By taking the derivative of the first
Einstein equation and replacing $\dot H$ from the second one, we find
\begin{equation}
  0 = 3H \dot \phi^2 + \ddot \phi \dot \phi + \dot V = (\ddot \phi +
  3H \dot \phi + \frac{\delta V}{\delta \phi} ) \dot \phi \; .
\end{equation}

Constant roll was defined in \cite{MS1} by asking that the ratio
between the acceleration and the speed of the field $\phi$ is
proportional to the Hubble constant
\begin{equation}
  \label{cr1f}
  \ddot \phi = - \eta H \dot \phi \; .
\end{equation}
Here we shall adopt an equivalent formulation in terms of the Hubble
parameter only
\begin{equation}
  \label{ctroll}
  \ddot H = - 2 \eta H \dot H \; ,
\end{equation}
and it is easy to see from the Einstein equations that the above
definitions are indeed equivalent. This latter definition is more
fundamental in the sense that it is enough to determine the
inflationary dynamics independent on the field content of the
theory. We shall use this property when we generalise constant roll to
a matter system of two scalar fields.

Equation \eqref{ctroll} can be integrated analytically and one can
obtain the solution for the Hubble constant and for the scale factor
as functions of time 
\begin{equation}
  \label{Hcr}
  \begin{aligned}
    H & = h \frac{k e^{h \eta t} + e^{- h \eta t}}{k    
      e^{h \eta t} - e^{- h \eta t}} \; , \\
    a & = C_a (k e^{h \eta t} - e^{- h \eta t})^{\frac{1}{\eta}} \; ,
  \end{aligned}
\end{equation}
where $h$ and $k$ are constants. This is the form of the solution
found in \cite{ASW} where the constants $k$ and $h$ are allowed to
take complex values as long as the solutions for $H$ and $a$ are
real. This reality condition imposes constraints on the way these
constants can be chosen and the various cases lead to the classes of
solutions discussed in \cite{MS1,ASW}. For simplicity we shall use
this general form as we are not particularly interested in a certain
class of models and bare in mind the fact that this formulation
encodes few specific possibilities depending on the reality of the
parameters $k$ and $h$.  Finally let us stress that under certain
conditions for the integration parameters and for the constant roll
parameter $\eta$ it was shown that inflation can take place and a
sufficient number of e-foldings can be accommodated
\cite{MS1,ASW}. Regarding the CMB fingerprints of inflation, like the
scalar and tensor perturbations, various studies in the literature
seem to favor small-$\eta$, but there is no true consessum on its
precise value \cite{MS2,ASW,GZF}.

So far we displayed the gravitational part of the solution which, up to this
point, can be seen to be independent on the matter content of the
theory. Now we need to make the link with the matter part. Note that
as long as we look for the potential as a function of time only, this
is given as
\begin{equation}
  \label{Vg}
  V = 3H^2 + \dot H \; ,
\end{equation}
which, after replacing the solution for the Hubble parameter above,
becomes
\begin{equation}
  \label{Vt}
  V(t) = 3h^2 + \frac{4kh^2}{(k  e^{h \eta t} - e^{- h \eta t})^2} (3- \eta) \; .
\end{equation}
Furthermore, the time profile for the scalar field can be obtained
using the second Einstein equation and the solution for the Hubble
constant \eqref{Hcr}
\begin{equation}
  \label{phit}
  \phi(t) = \pm \sqrt{\frac{2}\eta} \ln \frac{\sqrt k e^{\frac12 h \eta
      t} - e^{- \frac12 h \eta t}}{\sqrt k e^{\frac12 h \eta t} + e^{-
      \frac12 h \eta t}} + \phi_0 \; ,
\end{equation}
where $\phi_0$ is an additional integration constant.
Finally, inverting the equation above and inserting the solution in \eqref{Vt}
one finds the general scalar potential which is suitable for constant
roll inflation with a single scalar field
\begin{equation}
  \label{V1f}
  V = \frac{h^2}{2}(3+ \eta)+ \frac{h^2}{2}
  \cosh{(\sqrt{2 \eta}(\phi-\phi_0)) (3- \eta)} \; .
\end{equation}
We see that the potential which is compatible with constant roll
conditions is completely fixed. This potential is derived from the
Einstein equations of motion together with 
the constant roll condition and therefore is the unique potential
which admits such a constant roll regime. Moreover, as we saw
before, a solution of the Einstein equations is also a solution
of the scalar field equation. Therefore we are dealing with a completely
consistent solution of the Einstein and field equations which 
exhibits constant roll behavior.

\section{Constant roll inflation in string theory?}

It is natural to ask whether constant roll inflation can be
encountered in certain particle models. In particular, we shall ask this
question in the context of string/supergravity constructions. We do not
attempt to make an extensive survey of possible string potentials, but
we shall rather concentrate on a toy model which may be
derived from generic string
compactifications with fluxes. As we can see from \eqref{V1f}, the
potential which is needed for single field constant roll inflation is
quite rigid and apriori it is not at all clear that some arbitrary
string model can accommodate such a potential. We shall nevertheless
see that the model we picked up comes quite close to generate suitable
conditions for constant the roll regime to take place.

We shall consider a $N=1$ supergravity model coupled to complex field
whose real part can be thought of as the radius of the compactification
manifold in a would be string model. Therefore we write the K\"ahler
potential as 
\begin{equation}
  \label{Kpot}
  K = - 3 \log (T + \bar T) \; .
\end{equation}
The superpotential we consider is inspired from string
compactifications with fluxes in a regime of large volume \cite{GVW,dCGLM}
\begin{equation}
  \label{W}
  W= e_0 + i e T + m T^2 + i m_0 T^3 \; ,
\end{equation}
where $e_0$, $e$, $m$ and $m_0$ are constants which parametrize the
fluxes.

With these specifications we can compute the scalar potential by the
well-known $N=1$ formula
\begin{equation}
  \label{N1pot}
  V = e^K \left(D_T W D_{\bar T} \bar W g^{T \bar T} - 3 |W|^2 \right)
  \; ,
\end{equation}
where $D_T$ is the K\"ahler covariant derivative
\begin{equation}
  D_T = \partial_T  + \partial_T K \; ,
\end{equation}
and $g^{T \bar T}$ is the inverse K\"ahler metric. Parametrising the
complex scalar field $T$ as
\begin{equation}
  T = \rho + ia \; ,
\end{equation}
we find
\begin{equation}
  \label{pot}
  V = - \frac1{\rho} ( e_0 m + \frac13 e^2)  + \frac{a}{\rho} (3 e_0 m_0 -
  \frac13 e m) - \frac{\rho^2 + a^2}{\rho}(e m_0 + \frac13 m^2) \; .
\end{equation}

Our main interest in this section is to see whether this potential is
suitable for constant roll inflation. To check this, first note that
the scalar fields are not canonically normalised as the kinetic term
takes the form
\begin{equation}
  g_{T \bar T} \partial_\mu T \partial^\mu \bar T = \frac3{4 \rho^2} 
  (\partial_\mu \rho \partial^\mu \rho + \partial_\mu a \partial^\mu a) \; .
\end{equation}
Redefining $\sqrt{3/2} \ln{\rho} = \phi$, we find for the kinetic
terms
\begin{equation}
  \mathcal {L}_{kin} = - \frac12 \partial_\mu \phi \partial^\mu \phi -
  \frac34 e^{2\sqrt{\frac23} \phi} \partial_\mu a \partial^\mu a
\end{equation}
while the potential becomes
\begin{equation}
  \label{potnorm}
  \begin{aligned}
  V =\; & e^{-\sqrt{\frac23} \phi} \left[- ( e_0 m + \frac13 e^2)  + a
  (3 e_0 m_0 - \frac13 e m) - a^2(e m_0 + \frac13 m^2) \right] \\    
  & - e^{\sqrt{\frac23} \phi} (e m_0 + \frac13 m^2) \; . 
  \end{aligned}
\end{equation}

In order to match this potential with the one which is suitable for
a constant roll behavior we need to eliminate the field
$a$. Extremizing the potential with respect to $a$ and inserting the
solution back into the potential we find
\begin{equation}
  V = e^{-\sqrt{\frac23} \phi} \left[- ( e_0 m + \frac13 e^2) +
    \frac{(3 e_0 m_0 - \frac13 e m)^2}{4(e m_0 + \frac13 m^2)} \right] -
  e^{\sqrt{\frac23} \phi} (e m_0 + \frac13 m^2) \; . 
\end{equation}
It is interesting to note that this potential has the right
exponentials to combine into a hyperbolic cosine provided the
coefficients match in the right way and it is not inconceivable that
for a certain choice of flux parameters this actually happens. More
problematic is that the constant term which appears in the potential
\eqref{V1f} can not be reproduced in this simple model. It may be 
possible that certain modifications of the initial data of the problem
(eg various corrections to the K\"ahler and/or superpotential) do
lead to the presence of a constant term in the potential
without spoiling the general structure observed above. This however,
has to be checked on a case by case basis and will not concern us here
any longer. 

We only note that provided we succeed to identify the constant roll
potential along the lines mentioned above, the constant roll parameter
$\eta$ is going to be fixed at a value
\begin{equation}
  \eta = \frac13 \; .
\end{equation}
A first restriction for this parameter in order for
inflation to occur is that $\eta \le \frac12$ \cite{ASW}, which the
above value satisfies. However, as mentioned in the previous
section, smaller 
values for $\eta$ seem to be favored by observations \cite{GZF} while
clearly such a regime is not accessible in the simple string model
considered so far.

The weak point in the discussion above is that the string potential
generally depends on more scalar fields (in our case on two fields),
while the constant roll potential was derived for a single inflaton
field. Above, we reduced the string model to one with a single field
by extremizing the potential along one direction in order to decide
whether the model is suitable for constant roll behavior.  However, a
more meaningful approach would be to allow for the possibility that all
fields participate in the constant roll regime and let the theory
itself decide whether some of the fields should be spectators (ie are
going to be fixed at the extremal values of the potential). This will
be the purpose of the next section where we shall study the constant roll
behavior for a system comprising two scalar fields.

\section{Two field constant roll inflation}

In this section we want to analyse how the constant
roll setup of  \cite{MS1,ASW}, can be generalised to a
system of two real scalar fields. Having more fields one can choose
the system to be not minimally coupled and allow a non-trivial metric
on the scalar field space. Nevertheless, we shall not consider a
completely general metric, but based on the simple model discussed in
the previous section we write the matter action
\begin{equation}
  \label{matac}
  S_{matter} = \int \sqrt{-g} d^4 x \Big( - \frac12 \partial_\mu
  \phi \partial^\mu \phi - \frac12 e^{2b(\phi)} \partial_\mu
  \chi \partial^\mu \chi - V( \phi, \chi) \Big) \; ,
\end{equation}
where for the moment $b(\phi)$ is some arbitrary function of $\phi$,
but later on we shall specialise to the specific form which we used in
the example before.

For a FRW background, the Einstein and (scalar) field equations read
\begin{eqnarray}
  \label{Efe}
  3 H^2 & = & \frac12 \dot \phi^2 + \frac12 e^{2b} \dot \chi^2 + V \; ,
              \nonumber \\
  -2 \dot H & = & \dot \phi^2 + e^{2b} \dot \chi^2 \; , \\
  0 & = & \ddot \phi + 3H \dot \phi - e^{2b} \frac{\delta b}{\delta \phi} \dot
  \chi^2 + \frac{\delta V}{\delta \phi} \; , \nonumber \\
  0 & = & e^{2b} \big( \ddot \chi + 3H \dot \chi + 2 \dot \chi \dot b \big) +
  \frac{\delta V}{\delta \chi} \; , \nonumber
\end{eqnarray}
Compared to the single field case, the scalar field equations of
motion are no longer a consequence of the Einstein
equations. Following the same strategy by taking the time derivative
of the first Einstein equation and replacing $\dot H$ from the second
one we find
\begin{equation}
  \label{int}
  \dot \phi \left( \ddot \phi + 3H \dot \phi - e^{2b} \frac{\delta
      b}{\delta \phi} \dot \chi^2 + \frac{\delta V}{\delta
      \phi} \right) + 
  \dot \chi \left( \ddot \chi + 3H \dot \chi + 2 \dot \chi \dot b +
    \frac{\delta V}{\delta \chi} \right) =0 \; ,
\end{equation}
and we see that in the brackets we obtain precisely the field
equations. Therefore, if we find a solution of the Einstein equations
which satisfies one of the field equations, the other one is implied
by the relation above.  

As in the previous case, we are interested to see under what
conditions the above system admits 
constant roll solutions i.e. $\ddot H/ (H \dot H)  = 2 \eta =$ constant.
Fixing this relation for $H$ ensures that the solutions for the Hubble
parameter and the scale factor are going to be the same as in the
single field case \eqref{Hcr} and therefore, the conditions for
inflation are the same as the ones found in \cite{ASW}. However our
interest is to find the field profiles and the scalar potential which
gives rise to such constant roll inflation. 
We therefore need to find a suitable Ansatz for the
scalar fields such that this condition is satisfied. It is easy to see
that the equations of motion above require that
\begin{equation}
  \label{mfcr}
  \dot X = -2 \eta H X \; ,
\end{equation}
where $X= \dot \phi^2 + e^{2b} \dot \chi^2 $. This condition is the
two-field equivalent of \eqref{cr1f} and it reduces to the single field
condition if we set one of the fields to be constant.\footnote{
The fact that the kinetic term $X$ has to
satisfy the relation above is similar to multi-field slow 
roll inflation where precisely the same combination has to satisfy the
slow roll conditions.} 

Having the solution for $H$ from the single-field analysis
\eqref{Hcr}, we can find
$X$ immediately from the second Einstein equation as
\begin{equation}
  \label{Xsol}
  X = \frac{8 k h^2 \eta}{(k e^{h \eta t} - e^{- h \eta t})^2} \; . 
\end{equation}
A solution for $X$ in this case does not
immediately lead to a solution for $\dot \phi$ and $\dot \chi$ as we
saw in the single field case. To proceed further we shall make the
following Ansatz 
\begin{equation}
  \label{dotpc}
  \begin{aligned}
  \dot \phi & = 2h \sqrt{2 k \eta} \frac{\sin \theta}{k e^{h
                  \eta t} - e^{- h \eta t}} \; , \\
  \dot \chi & = 2h \sqrt{2 k \eta} \frac{\cos \theta \; e^{-b} }{k
                  e^{h \eta t} - e^{-h \eta t}} \; ,    
  \end{aligned}
\end{equation}
where $\theta$ is a free parameter. These are definitely sufficient
conditions for \eqref{Xsol} to be satisfied and there may be room for
generalizations which however will not be our concern in this
communication. Note that the parameter $\theta$ controls how much the
fields $\phi$ and $\chi$ participate in the constant roll regime. Indeed,
when either $\sin \theta$, or $\cos \theta$ vanishes one field will be
spectator and only the other one will participate in order to ensure
that the constant roll condition \eqref{mfcr} is satisfied.

For a more intuitive understanding of the Ansatz above
note that this is actually equivalent to require that both fields
$\phi$ and $\chi$ obey the corresponding one-field constant roll
condition \eqref{cr1f}, namely
\begin{equation}
\begin{aligned}
  \label{crcstr}
  \ddot \phi & =  -\eta H \dot \phi \; , \\
  \ddot \chi + \dot \chi  \dot b & = - \eta H \dot \chi \; ,
\end{aligned}
\end{equation}
where in the second equation we took into account the effect of the
non-trivial metric in the field space for $\chi$. In fact, once we
impose that the field $\phi$ obeys the 
constant roll condition \eqref{cr1f}, the condition on $\chi$ follows
automatically and then \eqref{dotpc} is the unique solution for $\dot
\phi$ and $\dot \chi$.

With these assumptions we are in position to find the solutions for
$\phi$ and $\chi$.  Up to the constant $\sin{\theta}$, the differential
equation for $\phi$ is the same as in the single field case and
therefore, up to this constant, the solution will be the same as in
Section 2
\begin{equation}
  \label{phisol}
  \phi(t) =  \frac{2 \sin \theta}{\sqrt{2\eta}} \ln \left( \frac{\sqrt k
      e^{\frac12 h \eta t} - e^{- \frac12 h \eta t}}{\sqrt k
      e^{\frac12 h \eta t} + e^{- \frac12 h \eta t}} \right) + \phi_0 \; .
\end{equation}

As long as $\chi$ is concerned, the non-trivial metric
on the field space becomes important. Without the factor $e^{-b}$ in
the expression for $\dot \chi$, the solution for this field would have
been the same as for $\phi$. The presence of this additional factor
changes the solution completely. At this stage we need a precise
definition for the function $b(\phi)$ and, as anticipated, we shall
choose the form used in the previous section when analysing the string
model
\begin{equation}
  b(\phi) = \sqrt{\frac23} \phi \; .
\end{equation}
Replacing this back into \eqref{dotpc} we find
\begin{equation}
  \label{dotchi}
  \dot \chi = 2h \sqrt{2 k \eta} \cos \theta \frac{\left( \sqrt k
      e^{\frac12 h \eta t} - e^{- \frac12 
        h \eta t} \right )^{2 \sqrt{\frac1{3 \eta}} \sin\theta - 1}}{ 
    \left( \sqrt k e^{\frac12 h \eta t} + e^{- \frac12 h \eta t} \right
    )^{2 \sqrt{\frac1{3 \eta}} \sin \theta + 1}} \; ,
\end{equation}
and this can be integrated to obtain
\begin{equation}
  \label{chisol}
  \chi(t) = \sqrt\frac32 \cot{\theta} \left( \frac{\sqrt k e^{\frac12 h
        \eta t} - e^{- \frac12 h \eta t}}{\sqrt k e^{\frac12 h \eta t}
      + e^{- \frac12 h \eta t}} \right)^{2 \sqrt{\frac1{3 \eta}} \sin
    \theta} + \chi_0 \; ,
\end{equation}
where again, $\chi_0$ is an integration constant.

At this stage we have obtained the time profile of the solution which
gives the constant roll behavior \eqref{ctroll} under the assumption
\eqref{crcstr}. Following the single field case, we need to
determine the scalar potential as a function of $\phi$ and $\chi$. As
mentioned before, when looking for the potential as a function of time,
the solution is given again by \eqref{Vg}, independent of the matter
content. While in the single field case, the field dependence of the 
potential was obtained simply by inverting the function $\phi(t)$ and
inserting it in \eqref{Vt}, when dealing with more scalar fields it is
not clear how to find the function $t(\phi, \chi)$ which should be
inserted in the potential. As we shall see, this ambiguity will remain
until the end and will constitute the main part for the flexibility of
the two-field potential.

Before that we have to make sure that the form of the potential is
consistent with the field equations of motion. In the single field case
this was automatic, but for two fields we need to explicitly impose
one of the field equations, the other one being implied by the
Einstein equations.
As mentioned before,
the solution in the two-field case depends on the parameter $\theta$
which quantifies the amount by which each of the fields contributes to
the constant roll regime. In the case $\theta = n \pi/2$ for some
integer $n$, one of the fields is a spectator and we are effectively
dealing with a single field regime. In these particular cases it is natural to
think that the individual potentials $V(\phi)$ and $V(\chi)$, obtained
by inverting the functions $\phi(t)$ and $\chi(t)$ and introducing
them into \eqref{Vt}, are the relevant scalar potentials.
After some straightforward algebra these potentials are found to be
\begin{eqnarray}
  \label{Vphi}
  && V(\phi) \equiv V(t(\phi)) = \frac{h^2}{2} (3 + \eta) +
  \frac{h^2}{2} (3-\eta) \cosh \left( \frac{\sqrt{2\eta}}{\sin
  \theta} (\phi - \phi_0) \right) \; , \nonumber \\ 
  & & V(\chi) \equiv V(t(\chi)) = \\[.5cm]
& & 3 h^2 + \frac14 h^2 (3- \eta) \left[
  \left(\frac{\chi - \chi_0}{\sqrt{\frac32} \cot \theta}
  \right)^{\frac{\sqrt{3 \eta}}{2 \sin \theta} } - \left(\frac{\chi -
  \chi_0}{\sqrt{\frac32} \cot\theta} 
  \right)^{-\frac{\sqrt{3 \eta}}{2 \sin \theta} }\right]^2 \;
  . \nonumber
\end{eqnarray}
The potential $V(\phi)$ has the same form as the
one found in the single field case \eqref{V1f} the only difference
being the the angle $\theta$ which enters in the argument of the
hyperbolic cosine.  For $\sin \theta=1$, the potential $V(\phi)$ is
precisely the one found in the single field case and since
$\cos \theta = 0$ equations \eqref{dotpc} imply that $\chi$ is
constant.  Moreover, the equations of motion tell us that for
$\dot \chi =0$ the potential is independent on $\chi$ and, as
anticipated, the potential in this case is simply given by $V(\phi)$.
In other words, for $\cos \theta=0$ and 
$\chi=$constant, the field profile \eqref{phisol} is a solution of the
equations of motion only if the full potential reduces to $V(\phi)$ in
\eqref{Vphi}. 

Likewise, for $\sin \theta =0$, the potential should
reduce to $V(\chi)$.\footnote{To take the limit $\sin \theta \to 0$ we
  should first note that in this case $\dot \phi =0$. This means that the
  warp factor $e^\phi$ is constant and so the differential equation
  for $\chi$ reduces to the one in the single field case. Therefore
  the solution in this case should be again the same with the solution in the
  single field case.} This means
that the appearance of the fields $\phi$ and $\chi$ in the potential
strongly depends on the angle $\theta$ and the full potential should
should interpolate between $V(\phi)$ for $\sin \theta =1$ and $V(\chi)$ for
$\cos \theta =1$. 
We therefore write the following Ansatz for the full potential
\begin{equation}
  \label{intpot}
  V(\phi,\chi) = (1-f(t)) V(\phi) + f(t) V(\chi) = V(\phi) - f(t)
  (V(\phi) - V(\chi)) \; ,
\end{equation}
where $f(t)$ is some arbitrary function of time which must assume the
value $f(t)=0$ for $\cos \theta =0$ and $f(t)=1$ for $\sin
\theta=0$. Note that when replacing $\phi$ and $\chi$ by the
corresponding solutions in \eqref{phisol} respectively \eqref{chisol},
$V(\phi) \to V(t)$ and $V(\chi) \to V(t)$ ensuring that the full
potential reduces to $V(t)$.
The appearance of the 
function $f(t)$ might be surprising, but as we will shortly see, we
need such a function so that we can solve the equations of
motion. In fact, the equations of motion will determine the form of
this function and then we shall still be left with the freedom to
choose $t$ as a function of either $\phi$ or $\chi$ or a combination
thereof. One may legitimately wonder whether for the potential
\eqref{intpot} equation \eqref{int} is still valid as in the
derivation it was mutually assumed that $V$ does not explicitly
depend on time. Note however, that since the function $f(t)$
multiplies the term  $V(\phi) - V(\chi)$ which identically vanishes
once we introduce the the solutions for the fields \eqref{phisol} and
\eqref{chisol}, term like $\dot f$ will not be present when taking the
time derivative of $V$ in \eqref{intpot}. For the same reason, even if we 
replace $t$ by a certain combination of $\phi$ and $\chi$, in the
equations of motion no derivatives of the function $f$ will appear.

With these in mind, we shall now impose the $\phi$ equation of motion
for the potential 
\eqref{intpot}. Replacing $\ddot \phi$ from the constant roll
condition \eqref{crcstr} and using the potential \eqref{intpot}, the
$\phi$ equation of motion becomes
\begin{equation}
  \label{phieq}
  - \eta H \dot \phi + 3H \dot \phi - e^{2b} \frac{\delta b}{\delta \phi} \dot
  \chi^2 - (1-f(t)) \frac{\delta V(\phi)}{\delta \phi} =0 \; .
\end{equation}
In this equation everything apart from the function $f$ is known and
replacing the solutions for $H$, $\dot \phi$, $\dot \chi$ and
$V(\phi)$ we find 
\begin{equation}
  \label{ft}
  f(t) = \cos^2 \theta \left[ 1 - \frac{4 \sqrt{\frac{\eta k}{3}} \sin
      \theta}{(3 - \eta) (k e^{h \eta t} + 
    e^{-h \eta t})} \right] \; .
\end{equation}
It is clear that this function satisfies our basic conditions that it
should take the value 1 for $\sin \theta =0$ and vanishes
for $\cos \theta =0$. As we anticipated the function has a
non-trivial $t$-dependence which is essential for the field equations
to be satisfied. Replacing back into the potential we find
\begin{equation}
  \label{finpot}
  V(\phi, \chi) = \sin^2 \theta V(\phi) + \cos^2 \theta V(\chi) +
  \frac{4 \sqrt{\frac{\eta  k}{3}} \sin \theta}{(3 - \eta)} 
  \frac{\cos^2 \theta \sin \theta}{(k e^{h \eta t} +
    e^{-h \eta t})} (V(\phi) - V(\chi)) \; .
\end{equation}
This is the two-field potential which is compatible with the constant
roll conditions \eqref{crcstr} in the sense that the solution of these
constraints \eqref{phisol} and \eqref{chisol} are also solutions of the
equations of motion corresponding to this potential. The time
parameter still enters this potential and should be replaced by some
function of $\phi$ and $\chi$. Imposing the equations of motion has
fixed the freedom of choosing the form of the function $f(t)$ in
\eqref{intpot} but we are still left with the arbitrariness of
choosing $t$ as a specific function of $\phi$ and $\chi$. Therefore
the scalar potential in the two-field case is no longer fixed and
it may be easier to tune a certain model to obtain a potential which
ensures the constant roll behavior. A more detailed analysis of the
potential \eqref{finpot} will be left for future work since an
immediate prescription of how to replace $t$ as a function of $\phi$
and $\chi$ is not clear. We shall nevertheless comment on the
form of the component potentials $V(\phi)$ and $V(\chi)$. 

As also mentioned before, $V(\phi)$ is pretty much the same as the
single field potential \eqref{V1f}. The main difference consists of
the fact that another free parameter, in the form of $\sin \theta$
enters the argument of the hyperbolic cosine. Recall that in the
simplistic string model we discussed before in the context of single
field constant roll, the constant roll
parameter was fixed at the value $1/3$ from the exponents obtained in
the potential \eqref{potnorm}. When comparing to the two-field case,
the constant roll parameter becomes
\begin{equation}
  \label{et}
  \eta = \frac13 \sin^2 \theta \; ,
\end{equation}
and therefore, an arbitrarily low value for $\eta$ can be obtained by
an appropriate choice of $\theta$ so that agreement with values
favored by observations can be easily reached.

In what concerns $V(\chi)$, note that it depends on $\chi$
power-like. This is encouraging as in the simple string model
discussed, the field $a$ also appears power-like in the
potential. With the identification of $\eta$ above, $V(\chi)$ becomes
\begin{equation}
  \label{Vchistring}
  V(\chi) =\frac32 h^2 + \frac1{12} h^2 \sin \theta + \frac18 h^2
  \left( 6 - \frac13 \sin^2 \theta  \right) \left( \frac{\chi - 
      \chi_0}{\sqrt{\frac32} \cot \theta}  - 
    \frac{\sqrt{\frac32} \cot \theta}{\chi - \chi_0} \right) \; .
\end{equation}
Finally, in this case the expressions for $\phi(t)$ and $\chi(t)$ can
be simplified 
\begin{eqnarray}
  \phi(t) = \sqrt{\frac32} \ln \left(\frac{\sqrt k e^{\frac12 h \eta
        t} -  e^{- \frac12 h \eta t}}{\sqrt k e^{\frac12 h \eta
        t} +  e^{- \frac12 h \eta t}} \right)^2 + \phi_0 \\
  \chi(t) =  \sqrt{\frac32} \cot \theta  \left(\frac{\sqrt k e^{\frac12 h \eta
        t} -  e^{- \frac12 h \eta t}}{\sqrt k e^{\frac12 h \eta
        t} +  e^{- \frac12 h \eta t}} \right)^2 + \chi_0 \nonumber
\end{eqnarray}
From this point it seems more tractable to find a replacement of $t$
in \eqref{finpot} in terms of $\phi$ and $\chi$, but we shall not
insist on this any further since it is meaningful only for a
specific model.

\section{Conclusions}

In this note we studied the conditions under which the constant roll
regime may appear in models with two scalar fields. Our
analysis was motivated in a first place by the fact that many
realistic models describe more than one scalar field. Moreover, the
one-field constant roll potential derived in \cite{MS1,ASW} is
completely fixed making it more difficult to match this potential with one
obtained from a specific model. 

Imposing a constant roll condition for each of the fields,
\eqref{dotpc}, we were able 
to derive the time profile of the fields and write the general form
for the potential. 
As expected, the two-field potential has a higher degree of complexity
and is no longer fixed, which makes it easier to use in specific models.
The main source for this flexibility is the fact that the time
parameter left in \eqref{finpot} has to be replaced by some function
of $\phi$ and $\chi$. The precise combination is quite arbitrary as
long as when replacing back the solutions \eqref{phisol} and
\eqref{chisol} the time parameter $t$ is recovered.

The second degree of flexibility, as compared to the single field
case, is represented by the parameter $\theta$ which
controls how much each of the fields $\phi$ and $\chi$ participate in
the constant roll regime.  In a
more indirect way, the same parameter $\theta$ dictates the amount by
which each of the individual potentials $V(\phi)$ and $V(\chi)$ enter
the full potential \eqref{finpot}.
For the string model presented, this
parameter is tied to the constant roll factor $\eta$ by the relation
\eqref{et}. This implies that, in the two-field case, by suitably
choosing $\theta$, the constant roll parameter can be made small
enough to match the preferred observational values.

Finally it should be mentioned that the solution discussed in the two
field case is still a particular one as the Ansatz \eqref{dotpc}
is not the unique solution for which relation \eqref{Xsol} holds.

\vspace{1cm}
\emph{Acknowledgments:} The author would like to thank D.~Ghilencea
for enlightening discussions at the beginning of this project. This
work was supported by project ``Nucleu'' PN-19060101/2019.


\begin{thebibliography}{}

%\cite{Candelas:1985en}
\bibitem{CHSW} 
  P.~Candelas, G.~T.~Horowitz, A.~Strominger and E.~Witten,
  ``Vacuum Configurations for Superstrings,''
  Nucl.\ Phys.\ B {\bf 258}, 46 (1985).
  doi:10.1016/0550-3213(85)90602-9
  %%CITATION = doi:10.1016/0550-3213(85)90602-9;%%


\bibitem{dWSHD} 
  B.~de Wit, D.~J.~Smit and N.~D.~Hari Dass,
  ``Residual Supersymmetry of Compactified D=10 Supergravity,''
  Nucl.\ Phys.\ B {\bf 283}, 165 (1987).
  doi:10.1016/0550-3213(87)90267-7
  %%CITATION = doi:10.1016/0550-3213(87)90267-7;%%


\bibitem{DPN} 
  M.~J.~Duff, B.~E.~W.~Nilsson and C.~N.~Pope,
  ``Kaluza-Klein Supergravity,''
  Phys.\ Rept.\  {\bf 130}, 1 (1986).
  doi:10.1016/0370-1573(86)90163-8
  %%CITATION = doi:10.1016/0370-1573(86)90163-8;%%


\bibitem{MN} 
  J.~M.~Maldacena and C.~Nunez,
  ``Supergravity description of field theories on curved manifolds and a no go theorem,''
  Int.\ J.\ Mod.\ Phys.\ A {\bf 16}, 822 (2001)
  doi:10.1142/S0217751X01003935, 10.1142/S0217751X01003937
  [hep-th/0007018].
  %%CITATION = doi:10.1142/S0217751X01003935, 10.1142/S0217751X01003937;%%


\bibitem{BCV} 
  T.~D.~Brennan, F.~Carta and C.~Vafa,
  ``The String Landscape, the Swampland, and the Missing Corner,''
  PoS TASI {\bf 2017}, 015 (2017)
  doi:10.22323/1.305.0015
  [arXiv:1711.00864 [hep-th]].
  %%CITATION = doi:10.22323/1.305.0015;%%


\bibitem{DvR} 
  U.~H.~Danielsson and T.~Van Riet,
  ``What if string theory has no de Sitter vacua?,''
  Int.\ J.\ Mod.\ Phys.\ D {\bf 27}, no. 12, 1830007 (2018)
  doi:10.1142/S0218271818300070
  [arXiv:1804.01120 [hep-th]].
  %%CITATION = doi:10.1142/S0218271818300070;%%


\bibitem{OOSV} 
  G.~Obied, H.~Ooguri, L.~Spodyneiko and C.~Vafa,
  ``De Sitter Space and the Swampland,''
  arXiv:1806.08362 [hep-th].
  %%CITATION = ARXIV:1806.08362;%%


\bibitem{AOSV} 
  P.~Agrawal, G.~Obied, P.~J.~Steinhardt and C.~Vafa,
  ``On the Cosmological Implications of the String Swampland,''
  Phys.\ Lett.\ B {\bf 784}, 271 (2018)
  doi:10.1016/j.physletb.2018.07.040
  [arXiv:1806.09718 [hep-th]].
  %%CITATION = doi:10.1016/j.physletb.2018.07.040;%%


\bibitem{Das} 
  S.~Das,
  ``A note on Single-field Inflation and the Swampland Criteria,''
  Phys.\ Rev.\ D {\bf 99}, no. 8, 083510 (2019)
  doi:10.1103/PhysRevD.99.083510
  [arXiv:1809.03962 [hep-th]].
  %%CITATION = doi:10.1103/PhysRevD.99.083510;%%

\bibitem{MS1} 
  H.~Motohashi, A.~A.~Starobinsky and J.~Yokoyama,
  ``Inflation with a constant rate of roll,''
  JCAP {\bf 1509}, 018 (2015)
  doi:10.1088/1475-7516/2015/09/018
  [arXiv:1411.5021 [astro-ph.CO]].
  %%CITATION = doi:10.1088/1475-7516/2015/09/018;%%


\bibitem{DRT} 
  M.~Dine, L.~Randall and S.~D.~Thomas,
  ``Supersymmetry breaking in the early universe,''
  Phys.\ Rev.\ Lett.\  {\bf 75}, 398 (1995)
  doi:10.1103/PhysRevLett.75.398
  [hep-ph/9503303].
  %%CITATION = doi:10.1103/PhysRevLett.75.398;%%



\bibitem{ASW}
  L.~Anguelova, P.~Suranyi and L.~C.~R.~Wijewardhana,
  ``Systematics of Constant Roll Inflation,''
  JCAP {\bf 1802} (2018) no.02,  004
  doi:10.1088/1475-7516/2018/02/004
  [arXiv:1710.06989 [hep-th]]


\bibitem{MS2}
  H.~Motohashi and A.~A.~Starobinsky,
  ``Constant-roll inflation: confrontation with recent observational data,''
  EPL {\bf 117}, no. 3, 39001 (2017)
  doi:10.1209/0295-5075/117/39001
  [arXiv:1702.05847 [astro-ph.CO]].
  %%CITATION = doi:10.1209/0295-5075/117/39001;%%


\bibitem{OO} 
  S.~D.~Odintsov and V.~K.~Oikonomou,
  ``Inflationary Dynamics with a Smooth Slow-Roll to Constant-Roll Era Transition,''
  JCAP {\bf 1704}, no. 04, 041 (2017)
  doi:10.1088/1475-7516/2017/04/041
  [arXiv:1703.02853 [gr-qc]].
  %%CITATION = doi:10.1088/1475-7516/2017/04/041;%%

\bibitem{NOO} 
  S.~Nojiri, S.~D.~Odintsov and V.~K.~Oikonomou,
  ``Constant-roll Inflation in $F(R)$ Gravity,''
  Class.\ Quant.\ Grav.\  {\bf 34}, no. 24, 245012 (2017)
  doi:10.1088/1361-6382/aa92a4
  [arXiv:1704.05945 [gr-qc]].
  %%CITATION = doi:10.1088/1361-6382/aa92a4;%%

\bibitem{MS3}
  H.~Motohashi and A.~A.~Starobinsky,
  ``$f(R)$ constant-roll inflation,''
  Eur.\ Phys.\ J.\ C {\bf 77} (2017) no.8,  538
  doi:10.1140/epjc/s10052-017-5109-x
  [arXiv:1704.08188 [astro-ph.CO]].

\bibitem{OOS} 
  S.~D.~Odintsov, V.~K.~Oikonomou and L.~Sebastiani,
  ``Unification of Constant-roll Inflation and Dark Energy with
  Logarithmic $R^2$-corrected and Exponential $F(R)$ Gravity,'' 
  Nucl.\ Phys.\ B {\bf 923}, 608 (2017)
  doi:10.1016/j.nuclphysb.2017.08.018
  [arXiv:1708.08346 [gr-qc]].
  %%CITATION = doi:10.1016/j.nuclphysb.2017.08.018;%%

\bibitem{CMP} 
  F.~Cicciarella, J.~Mabillard and M.~Pieroni,
  ``New perspectives on constant-roll inflation,''
  JCAP {\bf 1801}, no. 01, 024 (2018)
  doi:10.1088/1475-7516/2018/01/024
  [arXiv:1709.03527 [astro-ph.CO]].
  %%CITATION = doi:10.1088/1475-7516/2018/01/024;%%

\bibitem{GZF} 
  J.~T.~Galvez Ghersi, A.~Zucca and A.~V.~Frolov,
  ``Observational Constraints on Constant Roll Inflation,''
  arXiv:1808.01325 [astro-ph.CO].
  %%CITATION = ARXIV:1808.01325;%%


\bibitem{GVW} 
  S.~Gukov, C.~Vafa and E.~Witten,
  ``CFT's from Calabi-Yau four folds,''
  Nucl.\ Phys.\ B {\bf 584}, 69 (2000)
  Erratum: [Nucl.\ Phys.\ B {\bf 608}, 477 (2001)]
  doi:10.1016/S0550-3213(01)00289-9, 10.1016/S0550-3213(00)00373-4
  [hep-th/9906070].
  %%CITATION = doi:10.1016/S0550-3213(01)00289-9,

%\cite{dCGLM}
\bibitem{dCGLM} 
  B.~de Carlos, S.~Gurrieri, A.~Lukas and A.~Micu,
  ``Moduli stabilisation in heterotic string compactifications,''
  JHEP {\bf 0603}, 005 (2006)
  doi:10.1088/1126-6708/2006/03/005
  [hep-th/0507173].
  %%CITATION = doi:10.1088/1126-6708/2006/03/005;%%


\end{thebibliography}
\end{document}